\documentclass[conference]{IEEEtran}
\IEEEoverridecommandlockouts

\usepackage{cite}
\usepackage{amsmath,amssymb,amsfonts}
\usepackage{algorithmic}
\usepackage{graphicx}
\usepackage{textcomp}
\usepackage{xcolor}
\usepackage{url}
\usepackage{cleveref}
\def\BibTeX{{\rm B\kern-.05em{\sc i\kern-.025em b}\kern-.08em
    T\kern-.1667em\lower.7ex\hbox{E}\kern-.125emX}}
\usepackage{graphicx}
\usepackage{textcomp}    
\usepackage{xcolor}

\usepackage{tikz}
\newcommand*\circledB[1]{\tikz[baseline=(char.base)]{
            \node[shape=circle,fill,inner sep=0.2pt] (char) {\textcolor{white}{#1}};}}

\newcommand{\mysubsubsection}[1]{\noindent\textbf{#1}}
\usepackage[colorinlistoftodos]{todonotes} % For in-line todo comments in PDF 
\begin{document}
\IEEEoverridecommandlockouts
\IEEEpubid{\makebox[\columnwidth]{ 978-1-6654-1060-1/22/\$31.00 \copyright2022 IEEE \hfill} \hspace{\columnsep}\makebox[\columnwidth]{ }}

\title{Special Session: Towards an Agile Design Methodology for Efficient, Reliable, and Secure \\ML Systems\\
\thanks{*These authors contributed equally to this work.}
}

\author{\IEEEauthorblockN{
Shail Dave\textsuperscript{1*},
Alberto Marchisio\textsuperscript{2*},
Muhammad Abdullah Hanif\textsuperscript{3*}, 
Amira Guesmi\textsuperscript{4*},\\
Aviral Shrivastava\textsuperscript{1}, 
Ihsen Alouani\textsuperscript{4},
Muhammad Shafique\textsuperscript{3}
}

\textsuperscript{1}\textit{Arizona State University (ASU), USA}\\
\textsuperscript{2}\textit{Technische Universit{\"a}t Wien (TU Wien), Austria}\\
\textsuperscript{3}\textit{New York University Abu Dhabi (NYUAD), UAE}\\
\textsuperscript{4}\textit{Université Polytechnique Hauts-De-France (UPHF), France}\\
\\
shail.dave@asu.edu, 
alberto.marchisio@tuwien.ac.at,
mh6117@nyu.edu, 
amira.guesmi@uphf.fr \\
aviral.shrivastava@asu.edu, 
ihsen.alouani@uphf.fr,
muhammad.shafique@nyu.edu 
}

\maketitle

\begin{abstract}
The real-world use cases of Machine Learning (ML) have exploded over the past few years. However, the current computing infrastructure is insufficient to support all real-world applications and scenarios. Apart from high efficiency requirements, modern ML systems are expected to be highly reliable against hardware failures as well as secure against adversarial and IP stealing attacks. Privacy concerns are also becoming a first-order issue. This article summarizes the main challenges in agile development of efficient, reliable and secure ML systems, and then presents an outline of an agile design methodology to generate efficient, reliable and secure ML systems based on user-defined constraints and objectives.
\end{abstract}

\begin{IEEEkeywords}
ML, Neural Networks, Performance, Energy efficiency, DNN, Reliability, Security, Privacy, Agility, Robustness, Codesign.
\end{IEEEkeywords}

\section{Introduction}
The vast spectrum of real-world applications that can benefit from cutting-edge machine learning algorithms is driving the need for more efficient, dependable and secure ML systems. Moreover, the developments driven by this revolution are leading to new use cases and applications having more stringent constraints. For example, the recent developments in the domains of autonomous driving, robotics, smart healthcare, smart cities, and other always-on use cases are demanding to push the boundaries of energy efficiency, latency, reliability, security, accuracy, throughput, storage, and agility further to meet the ever-increasing real-world requirements. Conventional design methodologies and system stack tools are insufficient to support this revolution, and novel methodologies are required that can adapt to new workloads and efficiently generate the required system stack tools and techniques to build ultra-efficient and robust ML systems. Towards this, we highlight the following key challenges.

\textbf{Efficiency:} Hardware designers have conventionally defined Neural Processing Unit (NPU) accelerators via templates~\cite{venkatesan2019magnet, lai2020susy, zhang2022full}. An architectural template for an NPU specifies what kinds of computational and memory units can be interconnected and how. Various system stack tools for the NPU, such as cost models, simulators, and compilers, are developed manually by experts, limiting support to only the template architecture~\cite{zhang2022full, venkatesan2019magnet, dave2019dmazerunner}. As workloads evolve or application requirements become stringent, novel architectural features need to be integrated and explored~\cite{dave2021hardware}. But, tools from the prior system stack cannot be reused much without significant additional efforts for new architecture. Moreover, because design space is limited to exploring hyperparameters of one architecture, efficient architectures from the broad space remain unexplored, leading to inefficient designs. Further, existing accelerator design explorations~\cite{suda2016throughput, kao2020confuciux, zhang2022full, chen2018tvm, nardi2019practical} use black-box or NPU-agnostic optimizations; without reasoning about the effectiveness of explored solutions, they require thousands of trials or days for the vast space. 

\textbf{Reliability:} Conventionally hardware-induced reliability threats are mitigated through redundancy~\cite{vadlamani2010multicore, lyons1962use}, which, together with the compute-intensive nature of state-of-the-art Deep Learning (DL) models, translates to huge overheads. As the deep learning community is progressing towards deeper and more complex networks, it is imperative to consider cost-effective techniques designed by exploiting the intrinsic characteristic of DL models. A few cost-effective techniques have been proposed to address individual reliability issues~\cite{zhang2018analyzing, zhang2018thundervolt, chen2020ranger}; however, the literature still lacks systematic methodologies that can combine such techniques to offer an effective solution against the complete spectrum of reliability threats. 

\textbf{Security and Privacy:} Recent research works have highlighted several security and privacy issues~\cite{Goodfellow2015ExplainingAdversarialExamples,Shafique2020RobustMLDnT}, which make modern ML networks leak sensitive information and generate malicious results. Due to the variety of Deep Neural Network (DNN) models and vulnerability threats~\cite{Marchisio2019DeepLearningForEdgeComputing, Madry2018ResistentAdversarialAttacks, Shafique2021energyEfficientSecureEdgeAI}, common trends in the research community are to employ several security-oriented optimizations at different stages of the system design and implementation flow. However, it remains unclear how such security-oriented techniques can be integrated into a framework that combines other optimization objectives for ML systems, such as resiliency, latency, or energy efficiency.

\begin{figure}[t]
    \centering
    \includegraphics[width=\linewidth]{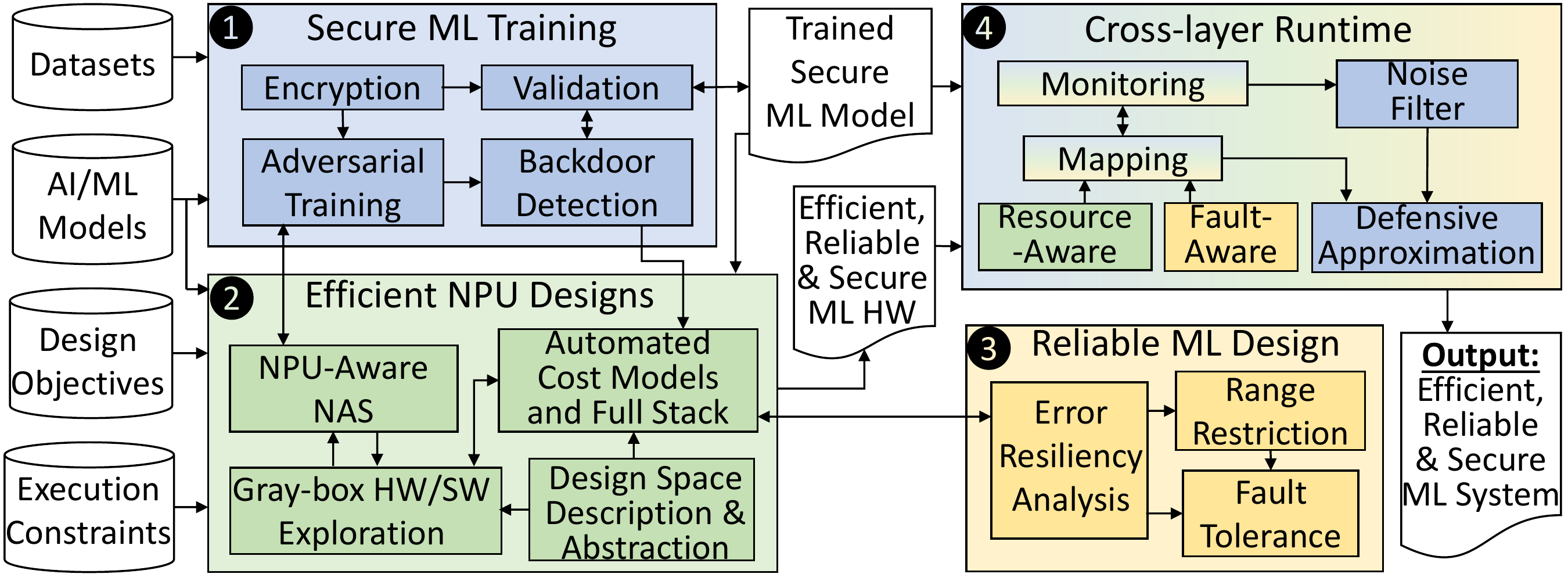}
    \caption{A cross-layer, agile methodology for designing efficient, reliable, and secure ML systems. Note that, in our approach, reliability and security are not mere after-thoughts but rather first-hand metrics in the codesign flow.}
    \label{fig:Introduction_figure}
\end{figure}

To address the aforementioned challenges, in this work, we present a methodology for designing and deploying efficient, reliable, and secure ML systems. The step-by-step flow of the proposed methodology is shown in \Cref{fig:Introduction_figure}.

\begin{enumerate}
    \item \textbf{Secure ML Training:} Given the ML models and datasets, security and privacy-preserving optimization techniques, such as adversarial training and backdoor detection are applied on encrypted data, to prevent privacy and adversarial vulnerabilities. Moreover, the resultant secure trained ML model is validated.
    \item \textbf{Efficient NPU Designs:} Based on design objectives and execution constraints for ML models, the methodology jointly explores the hardware and software configurations of various NPU architectures, as well as NPU-aware neural architectures. Tools for the stack for an NPU, including cost models, are automatically generated. The design space description and the cost models integrates the security and reliability metrics and analysis as well, enabling the design of efficient, reliable, and secure ML hardware.
    \item \textbf{Reliable ML Design:} The most advanced reliability optimizations are applied. Based on an error resiliency analysis, range restriction is employed to achieve high fault tolerant hardware.
    \item \textbf{Cross-Layer Runtime Techniques:} During execution, runtime monitoring is conducted to enable resource-aware and fault-aware mapping, and advanced security optimizations such as noise filtering and defensive approximation are applied. Consequently, the output is complete ML system that is efficient, reliable, and secure.
\end{enumerate}

Main contributions of this article are summarized as follows:
\begin{enumerate}
    \item We advocate for a cross-layer, agile methodology for designing efficient, reliable, and secure and privacy-preserving ML systems. Our approach builds upon the key idea that reliability and security should not be mere after-thoughts but rather first-hand metrics in a co-design flow.
    \item To mitigate efficiency-related challenges, we propose an agile design methodology. The key idea is to define design space description, which can enable comprehensive exploration of arbitrary architectures for target ML workloads. Our methodology uses bottleneck analysis for gray-box optimization, yielding explainability for obtained outputs and agile exploration.
    \item To mitigate reliability-related challenges, we highlight the most prominent cost-effective fault-mitigation techniques and then present a systematic methodology for effectively combining them. 
    \item To mitigate the security vulnerabilities and challenges, we discuss the most advanced privacy-preserving and security-preserving techniques for hardware-level intrusions and adversarial vulnerabilities, related to both the DNN-based conventional architectures and SNN-based neuromorphic architectures.
\end{enumerate}

\section{Agile Methodology for Designing \\ Efficient ML Systems}
The training and inference of ML models are increasingly done on domain-specific accelerators aka \textit{NPUs or Neural Processing Units}. Various applications impose stringent constraints on their execution on NPUs, which the design methodology for the NPU must meet. In this section, we review challenges faced by previous template-based design approaches. Then, we present an agile design methodology for obtaining efficient hardware/software codesigns for NPUs, along with automating full-stack development for a broad set of NPU architectures.  

\begin{figure}
\center {\includegraphics[width=\linewidth]{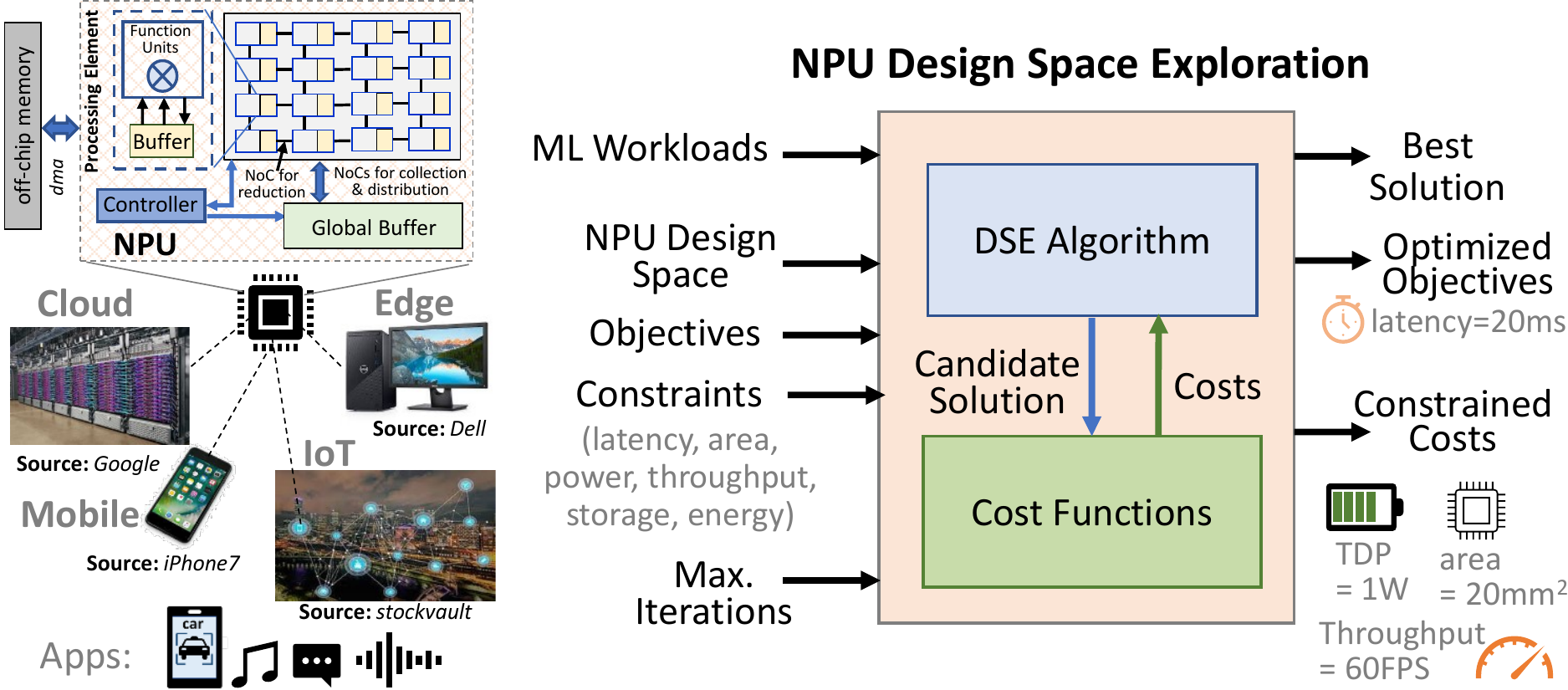}}
\caption{Exploring efficient hardware/software designs (right) for NPUs under different deployment scenarios (left).}
\label{fig:NPU-DSE}
\end{figure}

\begin{figure*}
\center {\includegraphics[width=0.8\linewidth]{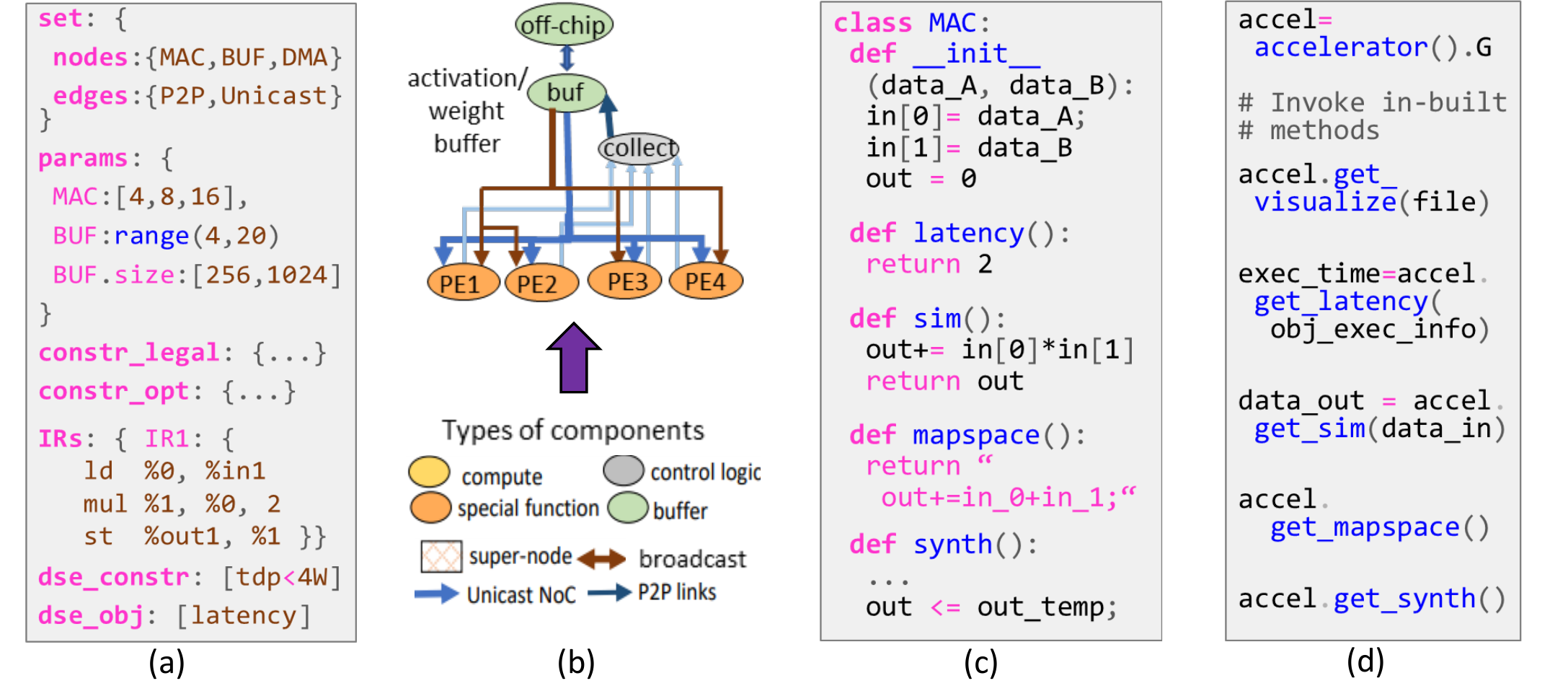}}
\caption{Examples for DSDL based design explorations. a) Design space specification. b) Obtaining a flow graph of an NPU during exploration. c) Design library for components. d) Design automation for a flow graph.}
\vspace{-0.15cm}
\label{fig:DSDL-overview}
\vspace{-0.15cm}
\end{figure*}

\subsection{NPU Design Requirements and Challenges}

\mysubsubsection{NPUs:} Examples of NPUs include Google's Tensor Processing Unit (TPU)~\cite{jouppi2017datacenter}, tensor cores in NVIDIA A100 Ampere architecture, Samsung NPU~\cite{park2021samsung}, Sambanova's RDU~\cite{emani2021accelerating}, IBM's AI Accelerator~\cite{fleischer2018scalable}, Microsoft Brainwave~\cite{chung2018serving}, Tesla's Self-Driving computer~\cite{talpes2020compute}, Facebook's ML accelerator~\cite{anderson2021first}, etc. NPU architectures can be standalone, a co-processor, or a near-data processing engine~\cite{kim2021aquabolt, gao2017tetris, song2019hypar}. Most NPUs are spatial architectures (e.g., Fig. \ref{fig:NPU-DSE}), while others like SpiNNaker and Intel Loihi~\cite{davies2021advancing} are neuromorphic processors. We present an agile, end-to-end methodology for obtaining efficient hardware/software designs of NPUs, with spatial architectures as a target example.

\mysubsubsection{NPU Design Flow Requirements:} Design space exploration (DSE) of NPUs require efficient HW/SW co-design that meets strict constraints of ML applications (Fig. \ref{fig:NPU-DSE}), i.e., within desired latency, area, power, throughput, accuracy, storage, energy budgets \cite{reddi2020mlperf}. Further, we need an agile design methodology because sustaining acceleration becomes challenging as ML workloads evolve. Besides, automatic and efficient construction of system stack is needed, as NPU architectures must adapt to new workloads by supporting specializations like sparsity or novel implementations such as mixed-precision computations~\cite{dave2021hardware}.

\mysubsubsection{Prior ADL-based Design Methods:} Prior NPU design approaches are mostly ADL-based, as in they define an \emph{architecture template}~\cite{boroumand2021google, minutoli2020soda, lai2020susy, mei2021zigzag, podobas2020template} in \textit{an architecture description language (ADL)}~\cite{chin2018architecture}, and build a system stack around it~\cite{zhang2022full}. \circledB{1} For a template, the architecture is fixed, i.e., what kind of computation and memory units are interconnected and how. The design space is limited to tuning only hyperparameters of this one architecture. It may lead to system \emph{inefficiency}, as a broad set of architectures is left unexplored. \circledB{2} Expert designers manually build tools like cost models, simulators, and compilers from scratch for the target architecture. This approach lacks flexibility. \circledB{3} Lastly, prior DSE approaches have used \emph{black-box} or \emph{NPU-agnostic optimizations} like simulated annealing or Bayesian~\cite{kao2020confuciux, nardi2019practical}, limiting the explainability and exploration agility. They cannot reason about high costs obtained or the effectiveness of potential design candidates. As a result, DSE can be costly in design time. The lack of dynamic DSE limits emerging applications such as deploying overlays for new DNN models on reconfigurable infrastructure or scheduling tasks of smart-city applications on distant, edge computing nodes, etc. 

\subsection{End-to-end Agile Design Workflow}

Our methodology addresses the aforementioned challenges via \textit{design space description language and framework (DSDL)}. The key idea is to allow comprehensive exploration of arbitrary architectures (Fig. \ref{fig:DSDL-overview}a), so that much efficient designs for target ML workloads can be explored. With flow graph abstraction of architectures (Fig. \ref{fig:DSDL-overview}b), DSDL can automatically build system stack tools like cost models, simulator, and compiler for an arbitrary flow graph (Fig. \ref{fig:DSDL-overview}d). Lastly, it uses bottleneck analysis which yields explainability and agile exploration.

\begin{figure*}
\center {\includegraphics[width=\linewidth]{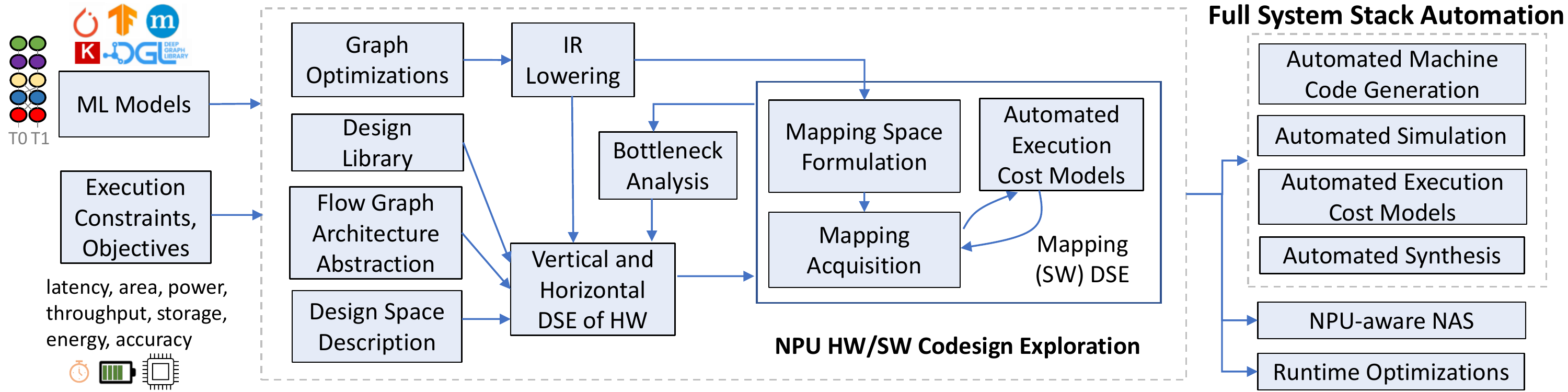}}
\caption{Agile methodology for designing efficient NPUs with automated full-stack development.}
\vspace{-0.1cm}
\label{fig:efficiency-workflow}
\vspace{-0.15cm}
\end{figure*}

\mysubsubsection{Overview of the Framework:} Fig. \ref{fig:efficiency-workflow} illustrates our agile design methodology for efficient ML accelerator designs. Application developers define ML models in frameworks like PyTorch, Tensorflow, or MXNet. ML compilers obtain graphs of these models and optimize them before lowering them into low-level IRs like LLVM. Developers also specify execution constraints and objectives for DSE. Our methodology enables comprehensive exploration by allowing designers and architects to specify the design space, i.e., components for computation, communication, memory, and control logic. Design libraries provide implementation of these components. Designers can also define rules for legality of the designs and pruning the search space, which are enforced during exploration of flow graphs (hardware architectures). Further, DSDL enables efficient codesign exploration through bottleneck analysis, which optimizes parameters that lower target costs. The functionality of various system stack tools (e.g., latency cost or programming the architecture) are auto-constructed by parsing the flow graph and aggregating the functionality corresponding to the individual nodes and edges. Since it allows determining cost models for arbitrary NPU architectures automatically, our methodology can yield efficient model/NPU codesigns through agile exploration. 

\mysubsubsection{Graph Optimizations:} Graph-level optimizations help improve computational and memory efficiency by applying various transformations. They include operator fusion, algebric simplification, strength reduction, etc.~\cite{li2020deep}. Existing ML compilers such as XLA, Glow~\cite{rotem2018glow}, TVM~\cite{chen2018tvm}, and nGraph provide such optimizations and lower graphs into intermediate representations (IRs) like LLVM IR~\cite{lattner2004llvm} or MLIR~\cite{lattner2020mlir}, which can be fed to design exploration mechanism.

\mysubsubsection{NPU Design Abstraction and Specification:} Design abstraction impacts the architectures that can be specified or explored by designers. Our methodology uses a flow graph representation~\cite{dave2022dsdl}. In an NPU's flow graph, nodes are primary components of computation, memory, control logic~\cite{pal2020transmuter, weng2020dsagen}, or even a sub graph and edges are fixed/reconfigurable interconnects. It allows modular construction of designs and modeling arbitrary hierarchy and grouping of arbitrary components. Moreover, it facilitates the support extension for novel architectures. Further, since algorithms are compiled as data flow graphs~\cite{das2018dataflow, dave2017ccf}, they can be conveniently mapped on flow graphs of NPUs. Flow graphs can be constructed through a library, which provides APIs for defining nodes and edges as well as their grouping and replication.   

\mysubsubsection{Libraries and Frameworks for Modular Accelerator Construction:} They define preliminary components for computation, memory, interconnections, and control~\cite{weng2020dsagen, wu2019accelergy}. The definitions specify simulation functionality, execution costs like area, latency, or energy~\cite{wu2019accelergy}, or hardware synthesis of each component~\cite{weng2020dsagen}. Frequently used high-level components can also be defined, such as vector units, systolic array, and computation/memory tiles~\cite{venkataramani2017scaledeep}, as specified in libraries like MAGNet~\cite{venkatesan2019magnet} and AutoDNNChip~\cite{xu2020autodnnchip}. Library also integrates reliability and security costs for components and specialized components such as razor flip flops for detecting timing violations or trusted memory. 

\mysubsubsection{Comprehensive NPU Design Space Formulation:} It determines NPU architectures and their designs that can be explored, and thus, overall efficiency. DSDL enables specifying vast space (Fig. \ref{fig:DSDL-overview}a), as hierarchy and organization of components are explicitly explored~\cite{dave2022dsdl}. Exploration of such broad space can lead to unseen architectures that are significantly more efficient (e.g., NPUs with no shared buffers for memory-bounded, no-reuse workloads~\cite{gao2017tetris}), unlike a specific architecture of the template. In DSDL, designers can define rules in terms of the legality of the generated designs, like what components can be interconnected or not and any relationships among their hyperparameters. In addition, there are also optimization rules for meaningful exploration in a pruned space. For instance, designers may want to explore homogeneous architectural components or a certain hierarchy of buffers. Designers also specify constraints, not just for latency, accuracy, etc., but also for reliability and security, based on such models supported by the NPU design library.

Based on the description, DSDL's iterative exploration takes place in three steps. The first is vertical exploration, which formulates a flow graph (Fig. \ref{fig:DSDL-overview}b) and ensures its compatibility with workloads. Then, horizontal exploration figures out optimized hyperparameters for the flow graph (e.g., buffer sizes). For each design, DSDL optimizes the software, e.g., mapping configurations, for it. In fact, all three steps can be jointly explored, especially through an explainable DSE.

\mysubsubsection{Automating Comprehensive Mapping Space Formulation:} Mapping space for an NPU encapsulates all schedules (aka iteration spaces in a polyhedral compiler~\cite{wolfe1989more, bondhugula2008practical}) that are possible corresponding to various loop optimizations like tiling, ordering, and unrolling, when executing a nested loop on an NPU~\cite{dave2019dmazerunner, li2020deep, dave2021hardware}. To develop a compiler for a customized NPU architecture, experts have previously formulated the mapping space manually~\cite{dave2019dmazerunner, venkatesan2019magnet, mei2021zigzag} or relied on NPU-agnostic loop optimizations~\cite{lattner2004llvm}. Then, compiler mapped operations for a schedule by software pipelining~\cite{weng2020dsagen, dave2018ramp}. In contrast, our methodology can automate formulation of the mapping space by deriving program representation for each component and for the overall flow graph~\cite{dave2022dsdl}. It helps deriving set of possible transformations for given functionality.    

\mysubsubsection{Automating Accurate Execution Cost Modeling:} For a flow graph, we can obtain accurate execution costs automatically (Fig. \ref{fig:DSDL-overview}d) by parsing the graph and aggregating costs of children (Fig. \ref{fig:DSDL-overview}c) for each parent node~\cite{dave2022dsdl}. Thus, it can not only model area or energy accurately (e.g., in Accelergy~\cite{wu2019accelergy}), but also total latency. Besides, for optimizing execution on commercial hardware, cost models may be obtained through machine/deep learning~\cite{adams2019learning} or profiling~\cite{wu2019fbnet}. Alongside, our method also allows incorporating security and reliability costs for components, including for data leaking and NBTI for aging, which helps formulating overall security and reliability models for the NPU design.

\mysubsubsection{Automated Machine Code Generation and NPU Simulation:} Design library provides simulation functionality for each component (Fig. \ref{fig:DSDL-overview}c), and optionally machine instructions to program the component (e.g., a PE). The collective simulation of an NPU architecture occurs as a dataflow through its flow graph. The control triggers that are necessary for a looped execution and live-in/out data are communicated via a synthetic controller, which also simulates non-accelerator functionality~\cite{dave2022dsdl}. The machine code can be generated by issuing instructions corresponding to a schedule.  

\mysubsubsection{Explainable HW/SW Codesign Exploration:}  Instead of black-box optimizations, our methodology uses bottleneck analysis for an explainable DSE~\cite{dave2022dsdl, yang2010automated}. Bottleneck analysis for flow graphs is obtained through constructing the cost graphs from cost models, indicating the costs consumed by various datapaths of the NPU architecture. It informs about factors incurring high execution costs for each candidate design explored (e.g., for inference latency, computation time is $n\times$ higher than the time for NoC communication, DMA transfers, or decoding compressed tensors~\cite{dave2020dmazerunner, dave2021hardware}) and values of the parameters that can mitigate them (e.g., number of PEs or function units). With explainability about obtained outputs, iterative DSE can yield quick convergence, enabling DSE of an NPU architecture at even run time! 

\mysubsubsection{Runtime Optimizations:} They ensure sustaining desired efficiency while meeting all necessary constraints, especially in distributed, multi-tenant~\cite{ghodrati2020planaria, quraishi2021survey} environments. Such optimizations include dynamic model selection from a model zoo~\cite{lou2021dynamic} (trade off accuracy with computations or memory footprint), data layout optimizations~\cite{li2020deep}, and DVFS (dynamic voltage-frequency scaling)~\cite{tu2020evolver}. Additionally, runtime can optimize execution by offloading computations of ML models partially to the cloud during processing on resource-critical embedded systems~\cite{almeida2021dyno,  ghasemi2021energy}. It may also coordinate contributions of participating devices for efficient convergence of federated learning models~\cite{ruan2021towards, kim2021autofl}. All such runtime optimizations can be incorporated with our framework, as they are either orthogonal or they could leverage obtained cost models for various NPUs and DSE for executing a model's layers onto individual devices.    

\mysubsubsection{NPU-aware NAS:} Recent techniques including~\cite{jiang2020standing, zhou2021rethinking, lin2021naas, choi2021dance, li2021flash} enable hardware-aware neural architecture search (NAS)~\cite{benmeziane2021comprehensive} of ML models. Our methodology empowers them to quickly explore efficient NPU/model codesigns, since they can leverage accurate cost models of arbitrary NPU architectures and efficient hardware/software explorations, while iterating through various ML model architectures for applications. 

\section{Designing Reliable ML Systems}
The success of DNNs has led to their adoption in safety-critical applications as well~\cite{lecun2015deep}\cite{sze2017efficient}. These applications, in general, include all such applications in which even a small error can lead to severe consequences. A few examples of these are autonomous driving, robotics, and healthcare analytics. As DNNs are highly computationally intensive, dedicated hardware accelerators are employed to offer energy-efficient data inference. These accelerators are usually built using modern nano-scale CMOS technology. However, the devices fabricated using such technologies are highly susceptible to different reliability threats such as soft errors, aging, and process variations, which may lead to errors and thereby catastrophic consequences. Moreover, these reliability threats are becoming an increasingly greater concern with technology scaling. 

Different mitigation techniques have been proposed to address reliability threats in modern systems. However, most of these techniques are based on redundancy. For example, Dual Modular Redundancy (DMR)~\cite{vadlamani2010multicore} and Triple Modular Redundancy (TMR)~\cite{lyons1962use} are two of the highly effective techniques for addressing different types of reliability threats, specifically soft errors. However, due to the compute-intensive nature of DNNs, these techniques result in ultra-high overheads and cannot be deployed in resource-constrained scenarios. Other techniques such as Error-Correcting Codes (ECC), critical variable re-computation, program duplication and instruction duplication~\cite{shafique2013exploiting} also have similar issues. \textit{Therefore, alternate techniques are required for DNNs to mitigate the effects of reliability threats without incurring huge overheads, ideally at ultra-low cost.} These techniques can be developed by exploiting intrinsic characteristics of DNNs, employing redundancy only for critical neurons/computation, or by transforming critical errors into non-critical ones through system modifications.

\subsection{Reliability Threats}

\begin{figure}[t!]
    \centering
    \includegraphics[width=\linewidth]{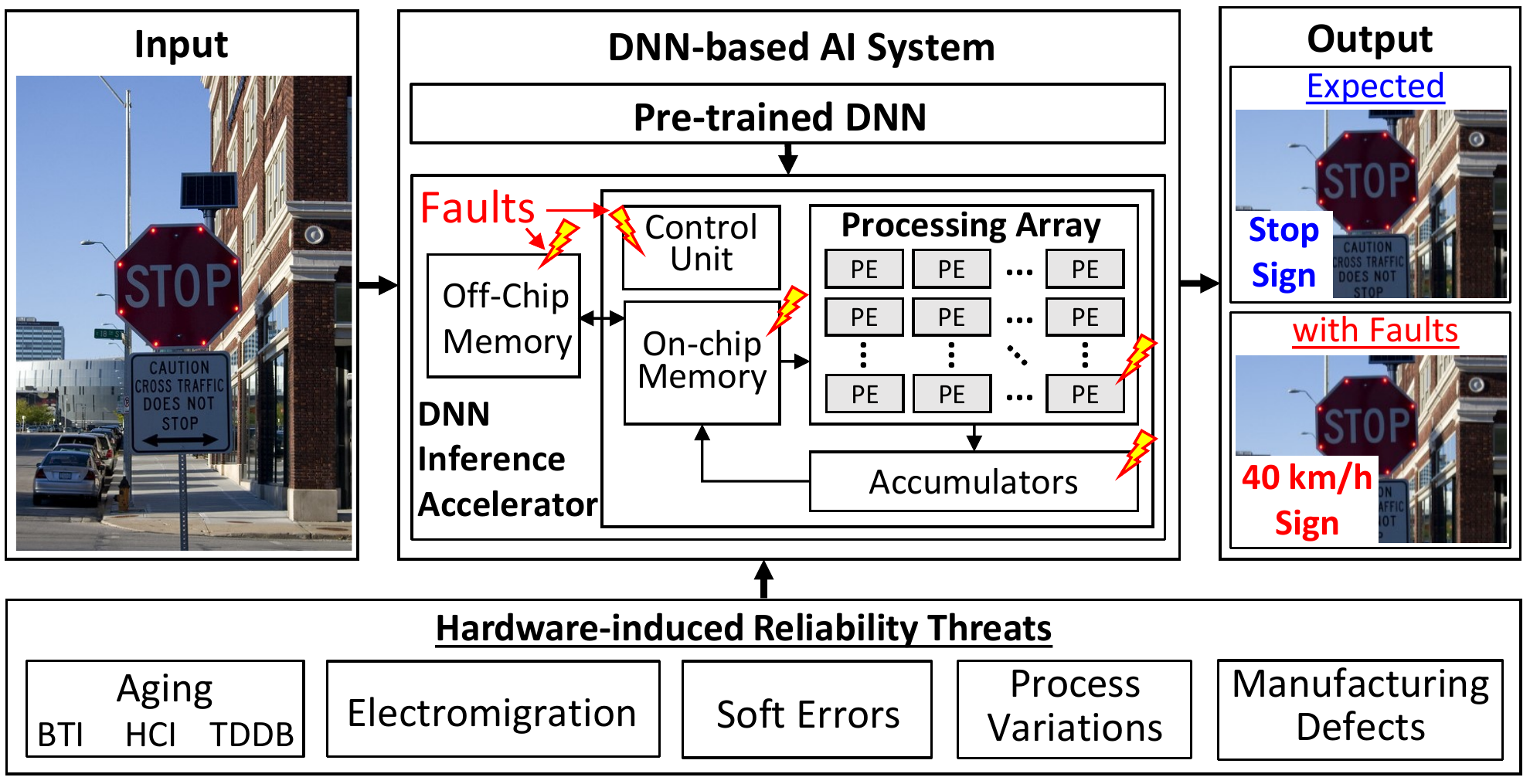}
    \caption{Reliability issues and their impact on the output of a DNN. (The stop sign image is from the COCO dataset~\cite{lin2014microsoft})}
    \label{fig:threats}
    \vspace{-0.1cm}
\end{figure}

Modern nano-scale devices face various reliability issues. Fig.~\ref{fig:threats} highlights the main types of reliability threats that can significantly degrade the performance of a DNN system. The figure also highlights a scenario in which these threats can lead to severe consequences, e.g., a fatal accident. The following text provides a brief introduction to different reliability threats. 

\begin{itemize}
    \item \textbf{Soft Errors} are bit-flips induced in computing systems due to radiation events. The main sources of these errors are alpha particles emitted by traces of impurities present in packaging materials and neutrons from cosmic radiation~\cite{baumann2005radiation}. These errors are transient in nature and vanish once new data is written to the faulty locations/cells. However, when present, they can significantly degrade the accuracy of an application. 
    
    \item \textbf{Aging} is gradual degradation of a fabricated device due to various physical phenomena like Bias Temperature Instability (BTI), Hot Carrier Injection (HCI), and Electromigration (EM). It mainly affects the hardware characteristics of circuits, for example, by increasing the threshold voltage ($V_{TH}$) of transistors~\cite{kang2008nbti} or by eroding the wires. Aging, in general, results in timing errors, which can also translate to permanent faults over time. 
    
    \item \textbf{Process Variations and Permanent Faults} are variations in the characteristics of transistors that occur during fabrication of an integrated circuit. These variations occur due to imperfections in the manufacturing process and affect the performance of the fabricated hardware. Process variations are, in general, addressed by adding guardbands, e.g., by increasing the supply voltage or decreasing the operating frequency of the chip/subsystem to ensure correct functionality. Extreme variations result in permanent faults and significantly affect the yield of the fabrication process. 
\end{itemize}

\begin{figure}[t]
    \centering
    \includegraphics[width=1\linewidth]{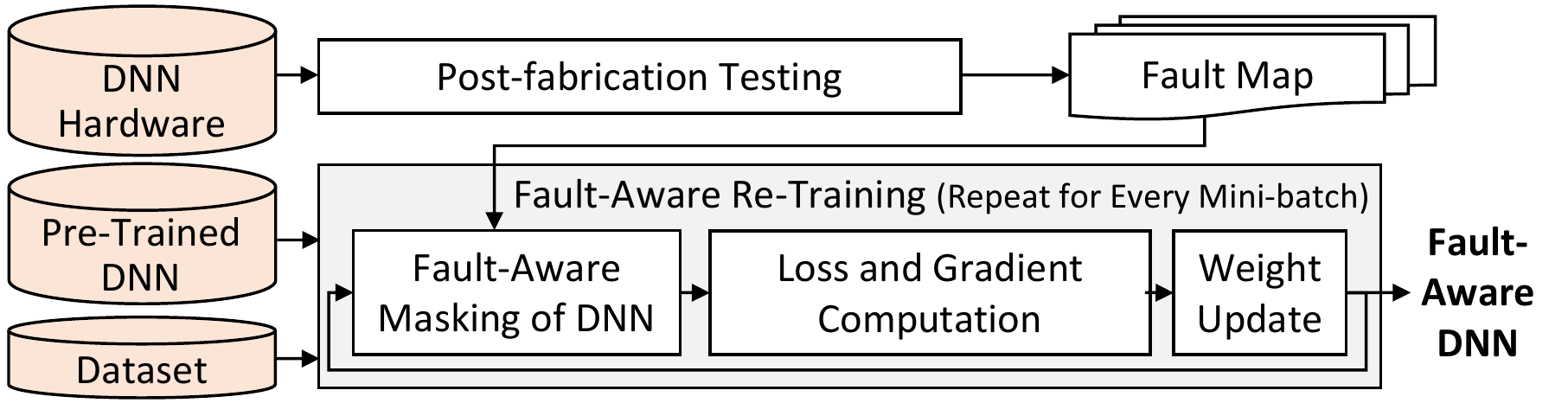}
	\caption{fault-aware re-training methodology}
    \label{fig:FA_retraining}
\end{figure}

\subsection{Cost-Effective Fault Mitigation}

Conventional redundancy-based solutions result in huge run-time and/or design-time overheads. To avoid such undesirable overhead costs, different low-cost techniques have been proposed in the literature to mitigated reliability threats in DNN systems. These techniques mainly exploit the intrinsic characteristics of DNNs or work on the principle of transforming critical faults to non-critical through low-cost modifications at the hardware as well as software level. The following text highlight the key concepts proposed for addressing different types of reliability threats in DNN systems without incurring huge overheads. 

\subsubsection{Soft Error Mitigation}

Robustness against soft errors is highly important, specifically for safety-critical applications and systems operating under harsh conditions. Studies have shown that these faults can result in significant degradation of the application-level accuracy when they occur at critical locations~\cite{hanif2018robust}\cite{chen2020ranger}. As conventional techniques result in huge overheads in DNN-based systems, specialized low-cost techniques are designed to improve the resilience of DNNs against soft errors. To address soft errors in on-chip memory, Azizimazreah et al. proposed a zero-biased SRAM cell design that has a higher probability of switching to '0' state in case an error occurs~\cite{8515692}. The design is based on the observation that 0-to-1 bit-flips in DNNs result in a higher accuracy drop compared to 1-to-0 bit-flips. To address soft errors in the computational array of DNN accelerators, Chen et al. proposed Ranger~\cite{chen2020ranger}, a range-restriction-based technique for improving the resilience of DNN systems against soft errors. It classifies all the activation values that fall out of a pre-defined range as faulty, and replaces them with a specific value from within the range. A similar technique is proposed in~\cite{Hoang2020FTClipAct} for mitigating soft errors in on-chip memory. Apart from the above-mentioned techniques, algorithm-based fault tolerance, such as checksum-based error detection and correction~\cite{Zhao2021AlgorithmBasedFaultTolerance}, and selective redundancy-based techniques, in which only the critical neurons are implemented redundantly, have also been proposed to mitigate soft errors at low cost. 

\begin{figure}[t]
    \centering
    \includegraphics[width=1\linewidth]{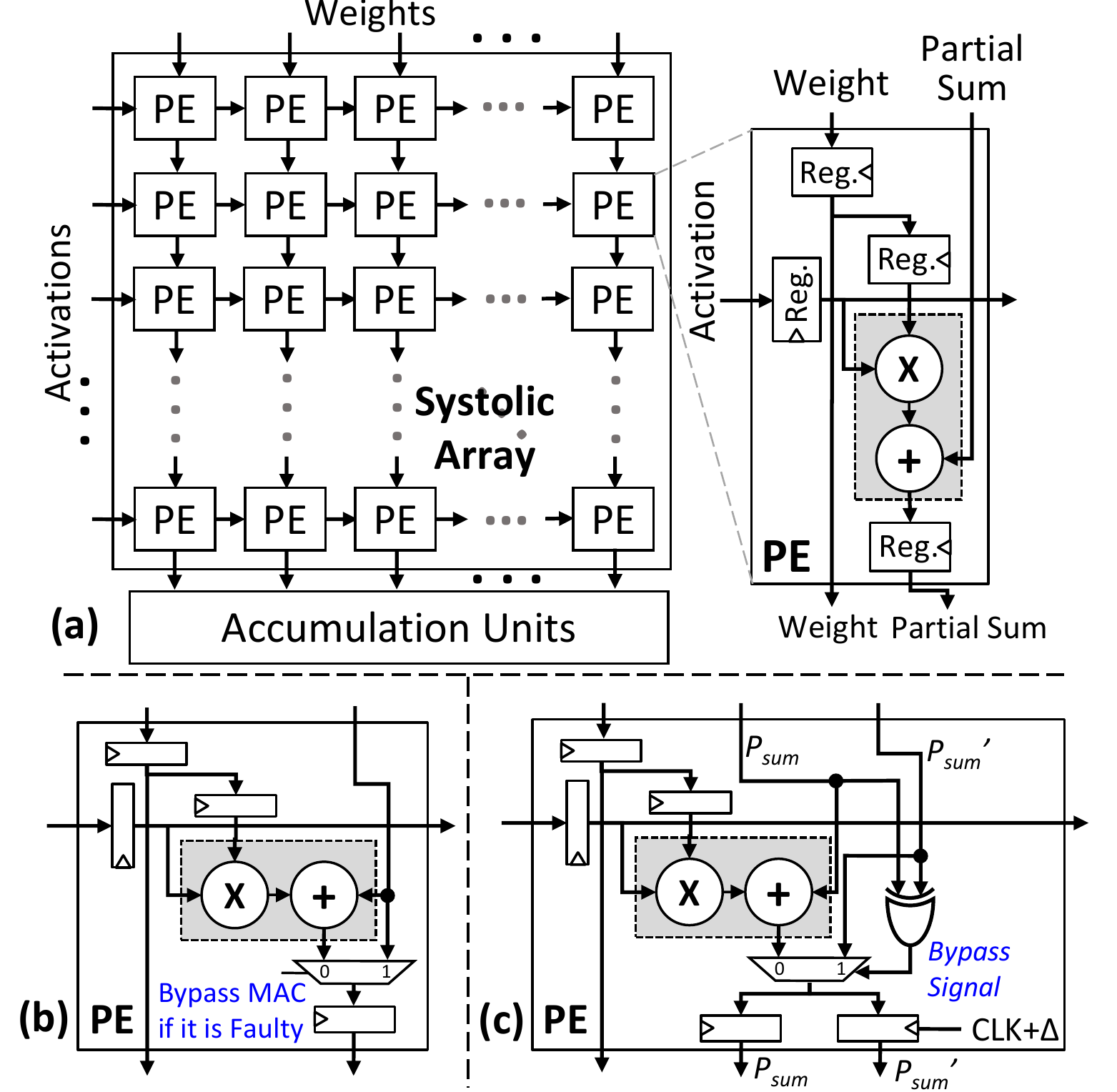}
	\caption{(a) A systolic array designed for accelerating DNN inference. (b) Modified PE design for for mitigating permanent faults~\cite{zhang2018analyzing}. (c) Modified PE design for mitigating timing errors~\cite{zhang2018thundervolt}.}
    \label{fig:DNN_Hardware}
\end{figure}

\subsubsection{Permanent Fault Mitigation}

Manufacturing defects or process variations induced permanent faults affect the yield of the chip fabrication process. The key goal of permanent fault mitigation techniques is to be able to improve the yield without incurring huge overheads and without affecting the performance of the systems in which the fabricated chips are deployed. As permanent faults are static in nature, the most effective technique for addressing them in DNN systems is Fault-Aware Training (FAT). To address permanent faults in the computational array of DNN accelerators, Zhang et al. proposed the concept of Fault-Aware Pruning (FAP)~\cite{zhang2018analyzing}. The technique exploits the fact that DNNs are, in general, resilient to small amount of pruning. To realize FAP, they added bypass paths in the processing elements of the array, which enable bypassing faulty MAC units. To further highlight the effectiveness of FAP, they coupled it with retraining using the methodology presented in Fig.~\ref{fig:FA_retraining} to achieve close to the baseline performance even at very high fault rates. Fig.~\ref{fig:DNN_Hardware}(b) shows the modifications required in the PEs of the array shown in Fig.~\ref{fig:DNN_Hardware}(a) to realize FAP approach. To address faults in the on-chip weight memory of DNN accelerators, Kim et al. proposed MATIC~\cite{kim2018matic}. It employs a methodology similar to Fig.~\ref{fig:FA_retraining}, however using memory fault maps, to generate a fault-aware DNN. Note that although the technique has been proposed mainly for improving the energy-efficiency of DNN systems through voltage scaling of on-chip weight memories, it can directly be employed for mitigating permanent faults in the on-chip memories.

\begin{figure}[t]
    \centering
    \includegraphics[width=1\linewidth]{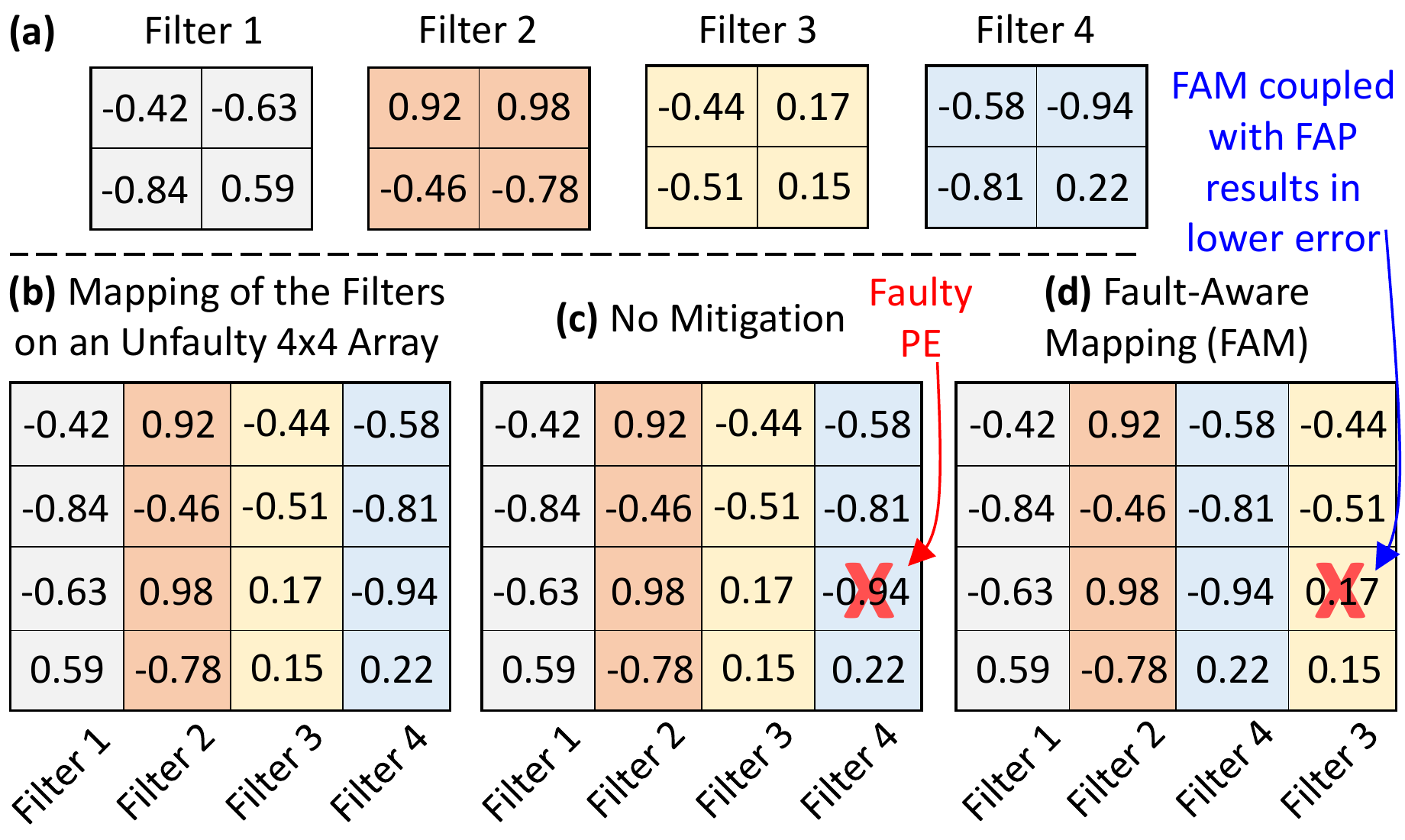}
	\caption{An example of fault-aware mapping technique using four filter and a 4x4 systolic array.}
    \label{fig:FAM_approach}
\end{figure}

Retraining DNNs requires huge computational resources. Therefore, FAT can result in huge design-time overheads, specifically in cases where a DNN has to be tuned for multiple faulty chip and each chip can have a distinct fault pattern. Under such circumstances, FAT cannot be employed. For such cases, Hanif et al. proposed SalvageDNN~\cite{abdullah2020salvagednn}, a saliency-driven fault-aware mapping technique. It works on the principle of mapping less significant weights on the faulty units. Fig.~\ref{fig:FAM_approach} presents an example of the working of SalvageDNN. 

\subsubsection{Aging Mitigation}

Aging in CMOS circuits affects the characteristics of transistors and manifests as timing errors. To detect and mitigate these errors in DNN accelerators, Zhang et al. proposed TE-Drop~\cite{zhang2018thundervolt}, a timing error recovery technique for systolic-arrays in DNN accelerators. The technique employs razor flip-flops for detecting timing errors in MAC units. Then, by exploiting the resilience of DNNs to pruning, mitigate each error by stealing a clock cycle from the corresponding downstream MAC unit and bypassing its update. Fig.~\ref{fig:DNN_Hardware}(c) presents the modifications required in the PEs of the array shown in Fig.~\ref{fig:DNN_Hardware}(a) to realize the concept. Pandey et al. proposed a similar concept, GreenTPU~\cite{8806805}, to mitigate timing errors by boosting the operating voltage of erroneous MAC units. Note that ThunderVolt and GreenTPU both have been proposed for boosting energy efficiency of a DNN system through voltage scaling. However, due to their effectiveness against timing errors, they are highly useful against aging in computational arrays of DNN accelerators as well. 

To alleviate aging of on-chip SRAM-based memory cells in DNN accelerators, Hanif et al. proposed DNN-Life~\cite{abdullah2021dnn}, a framework that employs read and write transducers to achieve ideal aging mitigation. The transducers mix data with random data, while keeping track of the encoding information, to balance the duty-cycle in each cell and minimize NBTI aging. 

\begin{figure}[t]
    \centering
    \includegraphics[width=\linewidth]{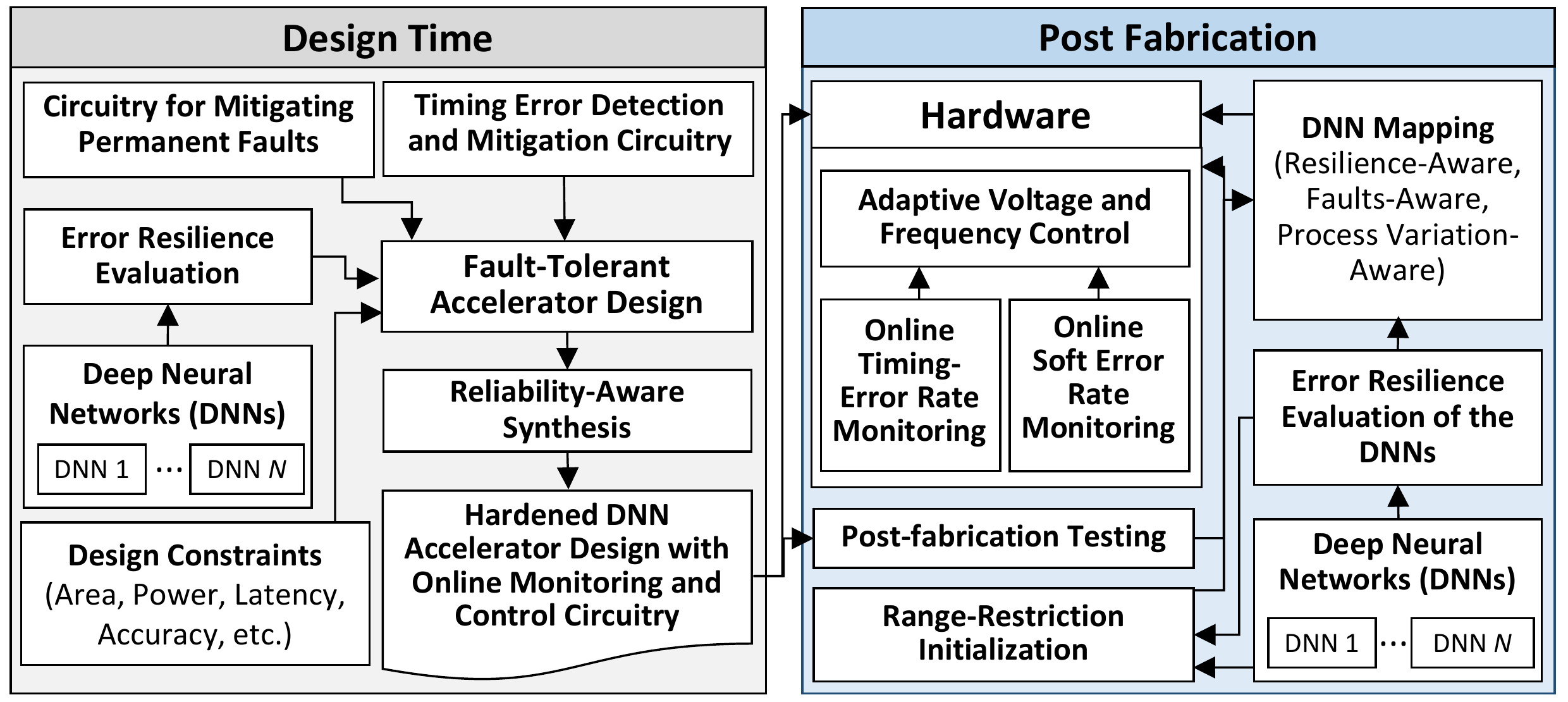}
	\caption{Overall Methodology for Building Reliable ML Systems~\cite{hanif2018robust}}
    \label{fig:Overall_meth_reliable}
\end{figure}

\subsubsection{Overall Methodology for Building Reliable ML Systems}

Fig.~\ref{fig:Overall_meth_reliable} shows our overall methodology for building reliable ML systems. In the design phase, a DNN accelerator is designed and then equipped with additional circuitry to support permanent fault mitigation and timing error detection and correction. The design is then passed through a reliability-aware synthesis~\cite{6464537} step, in which vulnerable nodes are selectively hardened to improve the system's resilience against soft errors. The hardware is then fabricated and passed for post-fabrication testing to locate permanent faults. The fault maps are then employed for fault-aware mapping and (if required) fault-aware re-training of DNNs. Once the DNNs have been tuned/adjusted, at the end, range-restriction methodologies are employed to define the bounds of activation values.

\section{Securing ML Systems}
\begin{figure*}[h!]
    \centering
    \includegraphics[width=\textwidth]{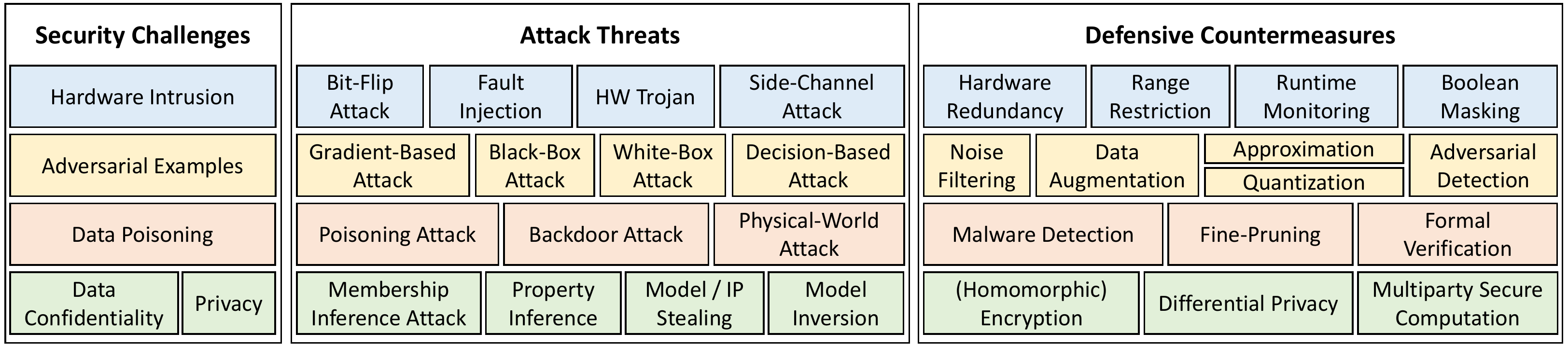}
    \caption{Overview of the Security threats for ML Systems.}
    \label{fig:MLSecurity_overview}
\end{figure*}

Since the DNN-based algorithms have achieved high accuracy for a large variety of tasks~\cite{Capra2020HWSWOptimizationsDNN, Capra2020UpdatedSurveyDNN}, it is legitimate to consider their employment in safety-critical applications~\cite{Shafique2021energyEfficientSecureEdgeAI, Neggaz2020CNNReliableCriticalApplications}. However, for these applications, it is crucial to guarantee robustness against security threats~\cite{Shafique2020RobustMLDnT, Neggaz2018ReliabilityCNN, Naseer2020FANNet, Marchisio2019CapsAttacks} to avoid catastrophic consequences. As shown in \Cref{fig:MLSecurity_overview}, such threats include vulnerabilities at the hardware-level, which undermine the correct functionality of computation and memory components, adversarial ML threats that aim at forcing a DNN to output wrong labels in the presence of corrupted inputs, and privacy-related issues. Due to the heterogeneity of the computational engines and their vulnerabilities, we discuss the security threats and their countermeasures for both conventional DNN architectures and SNN-based neuromorphic architectures.

\subsection{Hardware-Level Security Threats}

At hardware level, different types of vulnerabilities have been studied. When the DNN weights are stored in DRAM and SRAM memory cells, the bits can be flipped through Row-Hammer attacks~\cite{Kim2014RowHammer} or laser injection~\cite{Agoyan2010FlipBit}. While ML algorithms are relatively resilient to random bit-flips, the analysis conducted in NeuroAttack~\cite{Venceslai2020NeuroAttack} demonstrated that only 4 bit-flips in the most vulnerable weight locations are sufficient to fool DNNs and SNNs on the CIFAR10 dataset. The Bit-Flip Attack methodology~\cite{Rakin2019BitFlipAttack} uses a progressive bit search method to find the most vulnerable bits, while the work of~\cite{Bai2021FlippingWeightBits} proposed a method to generate targeted misclassification through bit-flips. More generically, fault-injection attacks~\cite{Liu2017FaultInjectionAttack} can target not only the weights, but also activation functions~\cite{Breier2018FaultAttack}. Moreover, the work of~\cite{Nagarajan2022FaultInjectionSNN} studied fault-injection threats for SNNs such as input spike corruption and SNN threshold manipulation.

The defensive techniques aiming at achieving \textbf{fault tolerance} are based on deploying hardware redundancy~\cite{Ozen2020BitErrorResilience}, or algorithm-based fault tolerance methods~\cite{Zhao2021AlgorithmBasedFaultTolerance} for detecting and correcting the errors in the convolutional layers. Concurrently, Ranger~\cite{chen2020ranger} directly rectifies the faulty output by applying a transformation that selectively restricts the value ranges in DNNs, and FT-ClipAct~\cite{Hoang2020FTClipAct} replaces the unbounded activation functions with their clipped versions, thus alleviating the impact of high-intensity faulty activation values.

Since hardware accelerator architectures are likely manufactured in off-shore fabrication facilities, they are vulnerable to \textbf{hardware trojans}, which are maliciously-introduced hardware modifications. The work of~\cite{Clements2018HWTrojanDesign} injected trojans into the computational engine to alter the behavior of DNNs' activation functions. NeuroAttack~\cite{Venceslai2020NeuroAttack} injects trojans into the weight memory which are triggered when a carefully-crafted adversarial pattern is recognized at the input of DNNs or SNNs. Since they can be detected through runtime monitoring~\cite{Khalid2018RuntimeTrojanMonitor}, a key feature for the hardware trojans to achieve stealthiness is having extremely low area and power consumption overhead, compared to the victim hardware design.

Moreover, power \textbf{side-channel attacks} can be applied to DNN hardware accelerators to recover the input image from the collected power traces, thus threatening their privacy integrity~\cite{Wei2018SideChannelAttackCNN}. The work of~\cite{Maji2021LeakyNets} studied power and time-based side-channel attacks for TinyML targeting embedded microcontrollers to recover the DNN model parameters and inputs. The state-of-the-art defensive methods against side-channel attacks are based on the Boolean masking~\cite{Dubey2021GuardingSideChannelAttacks}, in which all the hardware blocks for computing the linear and non-linear DNN operations are masked.

\subsection{Adversarial ML}

\noindent\textbf{Security for DNN-based Conventional Architectures:}
While ML applications are already in use in mainstream products and systems, they suffer from vulnerabilities to adversarial attacks that threaten their integrity and trustworthiness. In particular, adversarial examples modify an input to a ML classifier with carefully crafted perturbations chosen by a malicious actor to force the classifier to output a wrong label. If adversaries are able to manipulate the decisions of a ML classifier to their advantage, they can jeopardize the security and integrity of the system, and even threaten the safety of people it interacts with. For example, adding adversarial noise to a stop sign that leads an autonomous vehicle to wrongly classify it as a speed limit sign potentially leads to crashes and loss of life. In fact, adversarial examples have been shown effective in real-world context~~\cite{Eykholt2018PhysicalWorldAttacks, Man2020GhostImage}: that when printed out, an adversarially crafted image can fool the classifiers even under different lighting conditions and orientations. Therefore, understanding and mitigating these attacks is essential to developing safe and trustworthy intelligent systems.

%---------
%\subsubsection{Adversarial Attacks}
%---------
When attacking a DNN-based model, we can distinguish two main attack scenarios based on \textit{attacker knowledge}. A \textit{white-box setting} in which the adversary has complete knowledge of the training data of the victim model in addition to the target model’s architecture and parameters. In contrast, in a \textit{Black-box setting}, the adversary has partial or no access to the victim model’s architecture and parameters.  The adversary uses the results of querying the victim to reverse engineer the classifier and create a substitute model used to generate the adversarial examples. 
The attacker intention is to slightly modify the source image so that it is classified incorrectly by the target model, without special preference towards any particular output which is known as \textit{untargeted attack}. However, in a \textit{targeted attack}, the attacker aims at a specified wrong target class.
Attacks could be performed on different phases of the ML flow, and accordingly can be classified into two categories: \textit{Poisoning or training attacks}, when the attacker attempts to alter the training process by poisoning the training data in order to create specific classification errors~\cite{Shafahi2018PoisoningAttacks}. \textit{Backdoor attacks}~\cite{Gu2019BadNets} are also based on providing poisoned data to the victim to train the model with. In fact, the attacker aims to create a backdoor that allows the  input instances that are created using the backdoor key to be classified as a target label.
On the other hand, \textit{Inference attacks} or specifically \textit{evasion attacks} ~\cite{Madry2018ResistentAdversarialAttacks, MoosaviDezfooli2016DeepFool} are attacks that attempt to perturb an input in a way that it seems normal for a human but is wrongly classified by ML models.\\
Researchers have devised several defenses in response to these adversarial ML attacks, some of which focus on detection~\cite{Roth2019OddsDetection, Cohen2020DetectingAdversarialSamples}, while others focus on prevention and preparation of the ML model to defend against adversarial examples such as adversarial training~\cite{Madry2018ResistentAdversarialAttacks}, %where models are hardened by preemptively training the model on adversarial perturbations. This defense provides robustness against adversarial examples the classifier is trained on but any perturbation on which the classifier has not been trained can still evade the classifier. Another approach for hardening the ML models is based on 
input preprocessing~\cite{Khalid2019FAdeML, sit, das2017keeping},  %to increase the adversarial robustness. However, it was shown that this group of defenses is insecure under strong adaptive white-box attacks \cite{chen2019towards}.
Gradient masking based defenses~\cite{distillation_SP, nayebi2017biologically}. % relies on applying regularization to the model to make its output less sensitive to input perturbations. Papernot et al. proposed defensive distillation~\cite{distillation_SP}, which is based on increasing the generalization of the model by distilling knowledge out of a large model to train a compact model. Nonetheless, defensive distillation was found weak against the C\&W attack \cite{carlini2016evaluating}. Randomization-based Defenses rely on introducing random noise in the entire DNNs \cite{liu2018robust} or in some layers \cite{lecuyer2019certified}.
These defense strategies either alter the DNN structure, tweak the training procedure, or train the model only against known adversarial threats, which limits the defense scope to known vulnerabilities.
Another set of defense techniques are inspired by hardware-efficiency techniques such quantization~\cite{Lin2019DefensiveQuantization, Khalid2019QuSecNets}. Authors in~\cite{Guesmi2021DefensiveApproximation} proposed Defensive approximation (DA), which leverages approximate computing (AC) to build robust models. %In fact, defensive approximation \cite{Guesmi2021DefensiveApproximation} exploits a hardware-based data-dependent noise injected within model layers to help harden the DNN against adversarial attacks. The proposed defense requires no retraining, and reduces the DNNs resource utilization and energy consumption. proposed Defensive approximation (DA), which leverages approximate computing (AC) to build robust models. DA targets both robustness and energy/resource challenges; it uses a hardware-based AC-induced computational noise within the model layers to harden the DNN against adversarial attacks.
DA tackles the problem of robustness to adversarial attacks from a new perspective, i.e., approximation in the underlying hardware and targets both robustness and energy/resource challenges. In fact, DA exploits the inherent fault tolerance of deep learning systems~\cite{Neggaz2018ReliabilityCNN} to provide resilience while also obtaining by-product gains of AC in terms of energy and resources. The AC-induced perturbations tend to help the classifier generalize and enhances its confidence and consequently enhance the classifier's robustness.\\

\noindent\textbf{Security for SNN-based Neuromorphic Architectures:}
On neuromorphic architectures, the adversarial attacks and defenses can take advantage on different properties. For SNNs based on discrete data, white-box attacks~\cite{Bagheri2018AdvTrainingSNN} and black-box attacks~\cite{Marchisio2020IsSpikingSecure, Paudel2021SNNBlackboxAttacks} are generated and deployed. After generating the adversarial examples in the DNN domain, the work of~\cite{Sharmin2019ACA} demonstrated that the SNN generated through CNN-to-SNN conversion can be fooled by the same adversarial examples generated in the DNN domain. Moreover, the work of~\cite{Liang2020ExploringAttacksSNNGradient} proposed an attack algorithm based on the SNN gradient estimation both in the spatial and temporal domain.

Besides the conventional defense methodologies, recent work have demonstrated that it is possible to fine-tune the SNN structural parameters to improve its robustness. The work of~\cite{Sharmin2020SNNRobustness} studied the impact of discrete input encoding and non-linear activations, i.e., the leak factor in Leaky-Integrate-and-Fire (LIF) neurons, on the SNN's adversarial robustness. The work of~\cite{ElAllami2021SecuringSNNs} analyzed the SNNs robustness to adversarial attacks with different values of the LIF neuron's firing voltage thresholds and time window boundaries. Using these methods, the SNN robustness results up to $85\%$ higher than its equivalent DNN.

Since event-based sensing with dynamic vision sensors (DVS) is suitable for being deployed with high efficiency on low-power neuromorphic hardware, recent works demonstrated their applicability in safety-critical applications, such as autonomous driving, recognition and tracking~\cite{Viale_2021IJCNN_CarSNN, Massa2020EfficientSNN, Jiao2021EventCameraBasedSLAM}. Therefore, it is key to analyze the security aspects for event-based data. Towards this, the work of~\cite{Buchel2021AdversarialAttacksEventBased} modified the adversarial example generation algorithms of PGD, SparseFool, and Adversarial patches to be applied on event series. The work of~\cite{Lee2021AdversarialAttackEventBased} generated event-based adversarial examples on 3D point clouds. Moreover, the DVS-Attacks~\cite{Marchisio2021DVSAttacks} collect a set of stealthy yet efficient adversarial attack methodologies. The Sparse DVS-Attack injects sparse events in time, while the Frame, Corner, and Dash DVS-Attacks inject events in the whole time duration of the event sequences, injecting events in a frame around the sample, in a corner of the image, or in forms of dashed lines, respectively. Moreover, the Mask Filter-Aware Dash DVS-Attack is a modification of the Dash DVS-Attack, which limits the time duration of the perturbations.

\begin{figure}[t!]
    \centering
    \includegraphics[width=\linewidth]{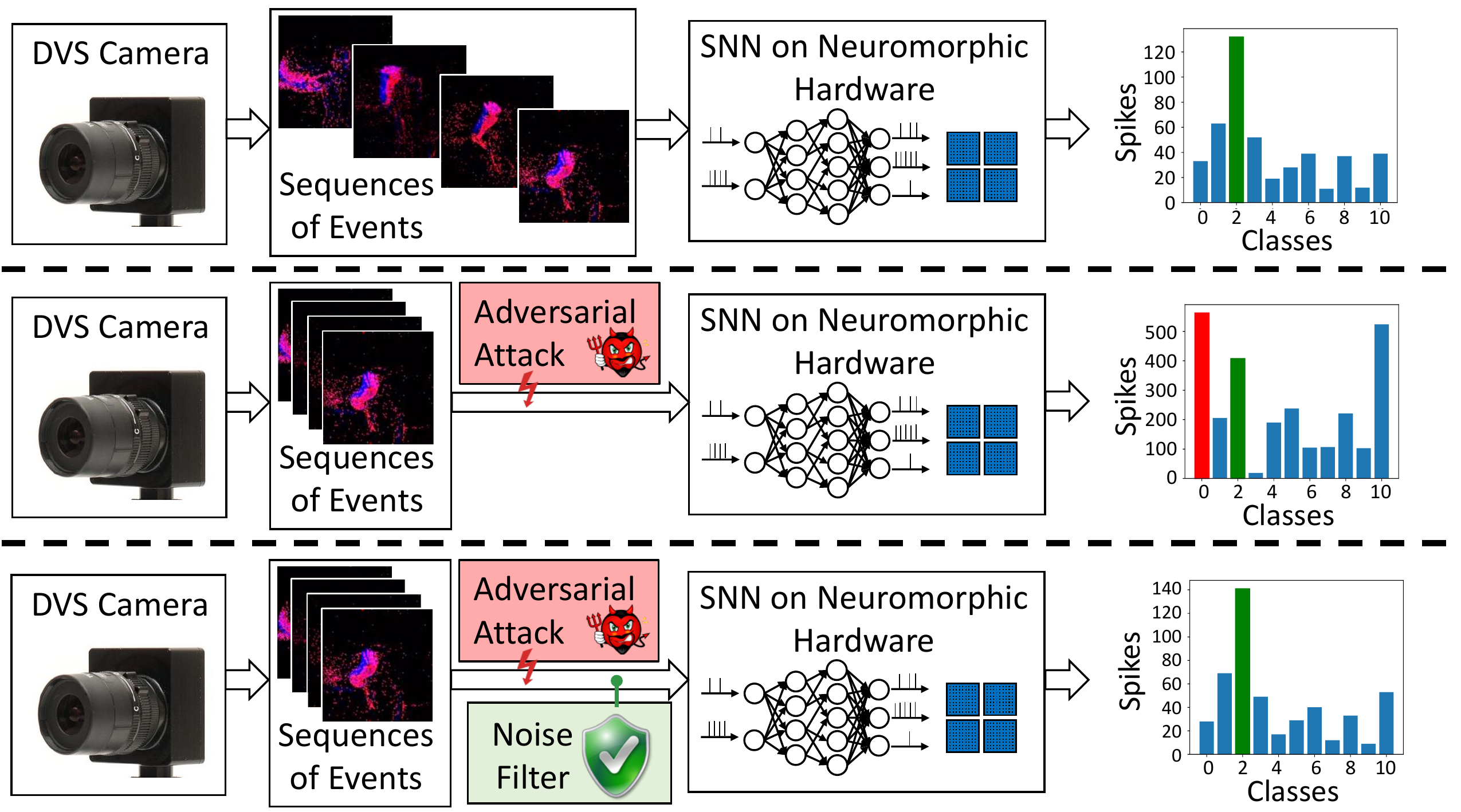}
    \caption{Example showing event-based adversarial attacks applied to gesture recognition systems implemented on neuromorphic hardware, in which the DVS-noise filter is applied to improve the SNN robustness~\cite{Marchisio2021RSNN}.}
    \label{fig:DVS_Attacks_Filter}
\end{figure}

While the noise filters for neuromorphic sensors~\cite{LinaresBarranco2019FilterDVS} have been originally designed for protecting against thermal noise and junction leakage fluctuation, their application to the input of neuromorphic computing engines serves as a defense mechanism against adversarial attacks~\cite{Marchisio2021RSNN} (see \Cref{fig:DVS_Attacks_Filter}). Among these \textbf{DVS-noise filters}, the background-activity filter (BAF) maintains only the events which have spatio-temporal correlation, thus filtering out the spurious events. The mask filter (MF) is designed to filter out the noise activity in pixels which have low temporal contrast. While the BAF makes the neuromorphic system robust against Sparse Attacks, the MF is resistant to Frame, Corner, and Dash Attacks. However, if the attacker has knowledge of the noise filter that is employed to protect the neuromorphic system, the MF-Aware Dash DVS-Attack~\cite{Marchisio2021DVSAttacks} can potentially break such defense.

\subsection{Privacy}

% Confidentiality is an explicit design property whereby one party wants to keep information (e.g., training data, testing data, model parameters, etc.) hidden from both the public and other parties (e.g., clients with respect to servers or vice-versa). Whereas, privacy is about protecting against unintended information leakage, whereby an adversary aims to infer sensitive information through some (intended) interaction with the victim.

ML algorithms are vulnerable to privacy threats, which are critical when data confidentiality is an issue, e.g., when revealing the identity of the patients in clinical records. Membership Inference Attacks~\cite{Shokri2017MembershipInferenceAttack} aim at determining whether a data sample belongs to the training dataset. More generically, Property Inference Attacks~\cite{Ganju2018PropertyInferenceAttacks} infer certain properties that hold only for a fraction of the training data, and are independent from the features that the DNN model aims to learn. On the other hand, Model Stealing methods~\cite{Tramer2016ModelStealing} aim at duplicating the functionality of the ML model and extract its parameters, and Model Inversion Attacks~\cite{Fredrikson2015ModelInversionAttack} aim to infer sensitive features of the training data.

Towards avoiding these leakages of confidential information, several privacy-preserving techniques can be employed. \textbf{Homomorphic Encryption (HE)} ensures that the data remains confidential, since the attacker does not have access to the decryption keys. CryptoNets~\cite{GiladBachrach2016CryptoNets} apply HE to perform DNN inference on encrypted data, and the work of~\cite{Nandakumar2019TrainingEncryptedData} extends the encryption to the complete training process. Another state-of-the-art technique is \textbf{Differential Privacy}, which can be guaranteed through the injection of noise to the stochastic gradient descent process (Noisy SGD)~\cite{Abadi2016DifferentialPrivacy}, or through Private Aggregation of Teacher Ensembles (PATE)~\cite{Papernot2018PATE}, in which the knowledge learned by an ensemble of ``teacher" models is transferred to a ``student" model. Several frameworks based on \textbf{Multiparty Secure Computation} have been designed, including SecureML~\cite{Mohassel2017SecureML}, Gazelle~\cite{Juvekar2018Gazelle}, and SecureNN~\cite{Wagh2019SecureNN}. Concurrently, Privacy-Preserving Domain Adaptation techniques~\cite{Kim2020PrivacyPreservingDomainAdaptation} preserve the privacy by transferring the knowledge from a labeled source domain to an unlabeled target domain. Moreover, in the PrivateSNN methodology~\cite{Kim2021PrivateSNN} the DNN-to-SNN conversion is followed by weight encryption with spike-based training on synthetic data for privacy-preserving SNNs.

\section{Conclusions}

Cutting-edge ML algorithms and applications are driving the need for more efficient, reliable, and secure ML systems, requiring a cross-layer codesign methodology. Key takeaways from the proposed methodology are:

\begin{itemize}
    \item Efficient designs for evolving ML workloads can be obtained by a design space description, which allows exploring various NPU architectures and design metrics, not just efficient designs of an NPU.
    \item Agile design flow can be enabled by full system stack automation for a wide range of NPU architectures and execution metrics.
    \item Agile design space exploration of ML systems can be enabled by reasoning about obtained costs and design decisions, e.g., with gray-box optimizations and bottleneck analysis.
    \item Reliability threats such as soft errors, process variations, permanent faults and aging can be mitigated in a cost-effective manner by exploiting intrinsic characteristics of DNNs. Moreover, a mitigation technique designed to address one threat can (to some extent) mitigate other reliability threats as well. Therefore, thorough exploration should be performed to select the best combination to collectively address all the threats. Moreover, as some of the techniques incur slight energy/power and area overheads, this exploration should be performed collectively with the search for an efficient NPU design. 
    \item Security threats like adversarial attacks, fault injections and privacy attacks can be detected by runtime fault monitoring or adversarial detection techniques; the robustness can be increased by applying adversarial training at design time, or noise filters and defensive approximation at runtime.
    \item Cross-layer codesign for efficiency, reliability, and security can be enabled by integrating corresponding cost models for design exploration, training of the models, and run-time monitoring and mapping.
\end{itemize}

\section*{Acknowledgment}

At ASU, this work was supported in part by NSF under Grant CCF 1723476—NSF/Intel Joint Research Center for Computer Assisted Programming for Heterogeneous Architectures (CAPA). 
At TU Wien, this work has been supported in part by Intel Corporation through Gift funding for the project ``Cost-Effective Dependability for Deep Neural Networks and Spiking Neural Networks'', and by the Doctoral College Resilient Embedded Systems, which is run jointly by the TU Wien's Faculty of Informatics and the UAS Technikum Wien. 
At NYUAD, different parts of these works were also supported in parts by the NYUAD Center for Interacting Urban Networks (CITIES), funded by Tamkeen under the NYUAD Research Institute Award CG001, Center for CyberSecurity (CCS), funded by Tamkeen under the NYUAD Research Institute Award G1104, and Center for Artificial Intelligence and Robotics (CAIR), funded by Tamkeen under the NYUAD Research Institute Award CG010.
At UPHF, this work has been supported in part by RESIST project funded by Région Hauts-de-France through STIMULE scheme (AR 21006614). 

\bibliographystyle{IEEEtran}
\bibliography{refsReliability, refsEfficiency, refsSecurity}

\end{document}